\documentclass[10pt,aps,prd,nofootinbib,superscriptaddress,twocolumn,preprintnumbers,balancelastpage]{revtex4}
\usepackage{amssymb,amsmath,latexsym,graphics, graphicx,epsfig,multirow,comment,hyperref,appendix}

\newcommand {\be} {\begin {equation}}
\newcommand {\ee} {\end {equation}}

\newcommand {\bes} {\begin {equation*}}
\newcommand {\ees} {\end {equation*}}

\newcommand{\es}[2] {\begin{equation} \label{#1} \begin{split} #2 \end{split} \end{equation}}

\newcommand{\beq}{\begin{equation}}
\newcommand{\eeq}{\end{equation}}

\newcommand {\kms} {\,\,\text{km}/\text{s}}

\newcommand{\V}[1]{{ \bf #1}}

\newcommand{\sun}{\odot}
\newcommand{\earth}{\oplus}
\newcommand{\Min}{\text{min}}

\newcommand{\esc}{\text{esc}}

\begin{document}
\title{Effect of Gravitational Focusing on Annual Modulation \\
in Dark-Matter Direct-Detection Experiments}
\author{Samuel K. Lee}
\affiliation{Princeton Center for Theoretical Science, Princeton University, Princeton, NJ 08544}
\author{Mariangela Lisanti}
\affiliation{Princeton Center for Theoretical Science, Princeton University, Princeton, NJ 08544}
\author{Annika H.~G.~Peter}
\affiliation{CCAPP and Departments of Physics and Astronomy, The Ohio State University, Columbus, OH 43210}
\author{Benjamin R. Safdi}
\affiliation{Department of Physics, Princeton University, Princeton, NJ 08544}

\date{\today}

\begin{abstract}
The scattering rate at dark-matter direct-detection experiments should modulate annually due to the Earth's orbit around the Sun.  The rate is typically thought to be extremized around \mbox{June 1}, when the relative velocity of the Earth with respect to the dark-matter wind is maximal.  We point out that gravitational focusing can alter this modulation phase.
Unbound dark-matter particles are focused by the Sun's gravitational potential, affecting their phase-space density in the lab frame.  Gravitational focusing can result in a significant overall shift in the annual-modulation phase, which is most relevant for dark matter with low scattering speeds.  The induced phase shift for light $O(10)$ GeV dark matter may also be significant, depending on the threshold energy of the experiment.  
\end{abstract}
\maketitle

An annually modulating signal at a direct-detection experiment is considered to be one of the tell-tale signatures of dark matter~\cite{Drukier:1986tm} (for a recent review, see~\cite{Freese:2012xd}).  Due to the motion of the Sun around the Galactic Center, there is a ``wind" of dark matter (DM) particles in the Solar reference frame. 
  This wind would result in a constant flux in the lab frame, but the Earth's orbit around the Sun leads, instead, to an annually modulating signal.

The time dependence in the detection rate can be seen explicitly as follows.  For typical spin-independent and -dependent interactions, the differential rate for a DM particle scattering off a target nucleus is proportional to 
\es{recoilR}{
{ d R \over d E_{\text{nr}} } \propto \rho \int_{v_{\text{min}}}^{\infty}\frac{ f\left( \V{v}, t\right)}{v}\,d^3v \, ,
}
where $\rho$ is the local DM density, $v_{\text{min}}$ is the minimum DM speed to induce a nuclear recoil with energy $E_{\text{nr}}$, and $f(\V{v}, t)$ is the DM velocity distribution in the lab frame~\cite{Smith:1988kw,Jungman:1995df}.  The time dependence in the rate is due to the changing distribution of DM velocities over a year.  

As explored in~\cite{Chang:2011eb,Lee:2013xxa}, a harmonic analysis of the modulation signal can lead to valuable information about the particle and astrophysics properties of the dark sector.  While~\cite{Lee:2013xxa} focused specifically on the contributions to the higher-order modes from the eccentricity of the Earth's orbit, the Galactic escape velocity, and velocity substructure, other physical effects can also come into play.  Here, we discuss focusing from the Sun's gravitational potential and its effects on the phase of the modulation.

The DM velocity distribution is warped by the gravitational field of the Sun, a phenomenon referred to as gravitational focusing (GF).  Specifically, the Sun's potential deflects the incoming, unbound DM particles, increasing their density and speed as they pass by the Sun.  The effect of GF on the interstellar medium around a star was considered by~\cite{Danby:1957, Danby:1967}, and the relevance of GF for DM was explored in~\cite{Griest:1987vc, Sikivie:2002bj, Alenazi:2006wu,Patla:2013vza}.  Ref.~\cite{Griest:1987vc} concluded that the effect on the total rate is negligible.
  In this Letter, however, we show that GF actually has a profound effect on the {\it phase} of the modulation and is highly relevant for current direct-detection experiments.
 
\begin{figure}[b]
\begin{center}
\includegraphics[width=3in]{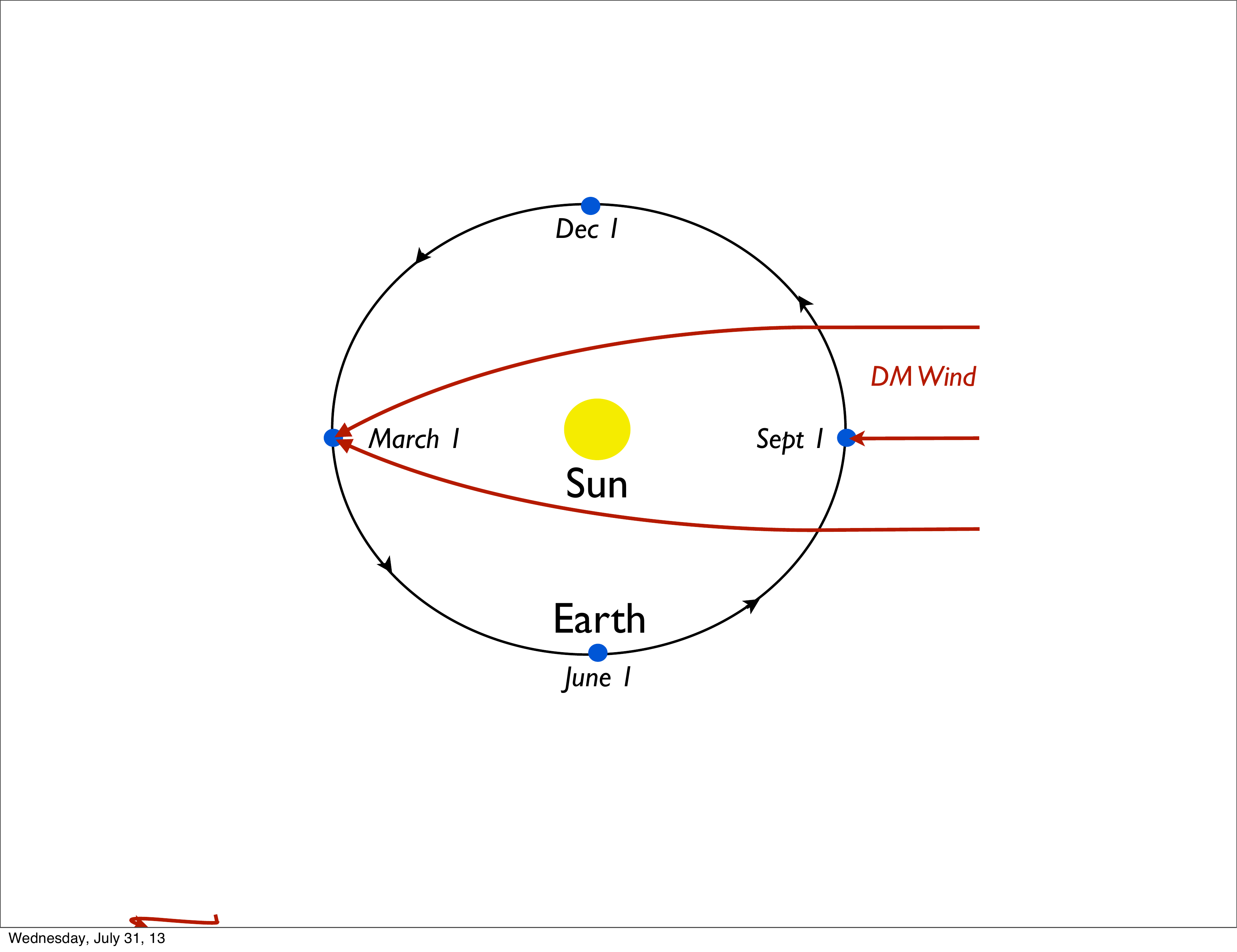}
\end{center}
\vspace{-.50cm}
\caption{A schematic illustration of the effect of gravitational focusing on unbound DM particles.  The phase-space density of DM at Earth is greater around March 1 than around September 1 due to this effect. }      
\vspace{-0.15in}  
\label{focusImage}
\end{figure}

GF affects the time dependence of the differential rate as follows.  The Earth is traveling fastest into the DM wind around June 1.  This means that during the fall \mbox{($\sim$September 1)}, the Earth is in front of the Sun, fully exposed to the DM wind, and during the spring \mbox{($\sim$March 1)}, it is behind the Sun.  As Fig.~\ref{focusImage} illustrates, GF is stronger during the spring than the fall because the DM particles have spent more time near the Sun; the changes in their density and velocity distribution are accordingly more significant.  Thus, when GF is accounted for, the time dependence in~\eqref{recoilR} arises not only from the velocity distribution but also from the density.  The effect on the rate is more pronounced for slower-moving particles that linger in the Sun's potential.

To more precisely calculate the effect of GF, we use the fact that the phase-space density of the DM along trajectories is constant in time due to Liouville's theorem~\cite{Griest:1987vc,Alenazi:2006wu}.  Specifically, 
\es{frelation}{
\rho \, f(\V{v},t) = \rho_\infty \, \tilde f \left( \V{v_\sun} + \V{v_\infty} \left[ \V{v} +  \V{V_\earth} \right]\right) \,,
 }
  where $\tilde f(\V{v})$ and $\rho_\infty$ are the velocity distribution (in the Galactic frame) and density, respectively, asymptotically far away from the gravitational well surrounding the Sun.  Note that the velocity distribution and density at the detector location, $f(\V{v},t)$ and $\rho$, respectively, are each time dependent when focusing is taken into account.  
$\V{V_\earth}$ is the time-dependent velocity of the Earth in the Solar reference frame~\cite{Lee:2013xxa}, and $\V{v_\sun} \approx (11 ,232 ,7 ) \kms$ is the velocity of the Sun in Galactic coordinates~\cite{Schoenrich:2009bx}.  
$\V{v_\infty}\left[\V{v_s}\right]$ is the velocity in the Solar reference frame that a DM particle must have at asymptotic infinity to have a Solar-frame velocity $\V{v_s}$. 
Using the conservation of the Laplace-Runge-Lenz vector, Ref.~\cite{Alenazi:2006wu} showed that 
\es{VinftyS}{
 \V{v_\infty}[\V{v_s}] &= {v_\infty^2\V{v_s} + v_\infty(G \, M_\sun / r_s) \V{\hat r_s} - v_\infty \V{v_s} (\V{v_s} \cdot \V{ \hat r_s}) \over v_\infty^2 + (G \, M_\sun / r_s)-v_\infty (\V{v_s} \cdot {\bf \hat r_s}) } \,, 
 } 
 with $\V{\hat r_s}$ the time-dependent unit vector that points from the Sun to the Earth~\cite{Lee:2013xxa} and $r_s$ the distance between the Sun and the Earth.
Energy conservation requires that \mbox{$v_\infty^2 = v^2 - 2 \, G \, M_\sun / r_s$}. \begin{figure}[tb]
\begin{center}
\includegraphics[width=3in]{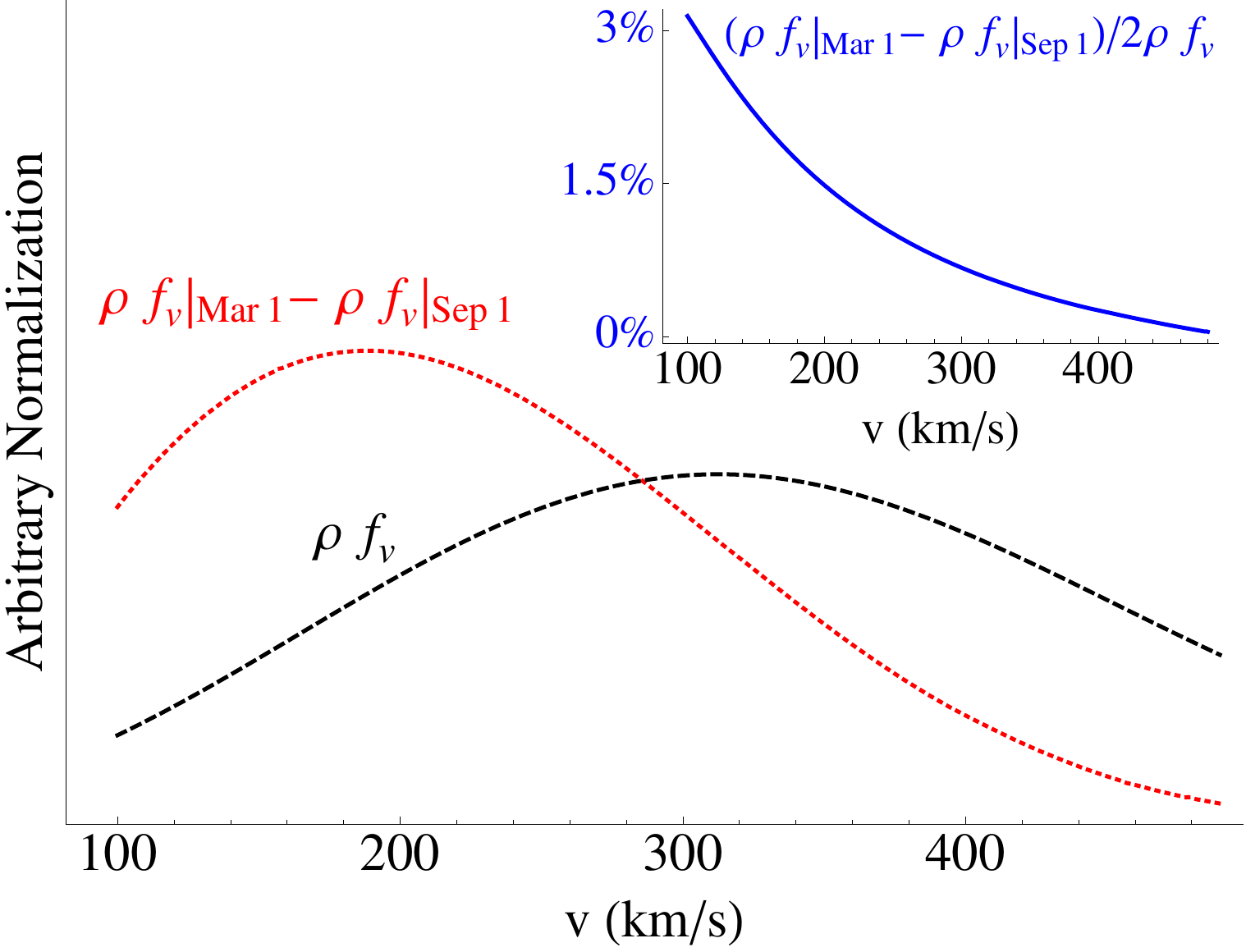}
\end{center}
\vspace{-.50cm}
\caption{For the standard halo model, the phase-space density $\rho f_v|_t$ (dashed black) is maximal at $300 \kms$; however, the difference in the phase-space density between spring and fall (dotted red) is maximal at $\sim$$200 \kms$.  Note that these two plots are normalized separately.  Gravitational focusing is 
most significant at low Earth-frame speeds $v$, as indicated by the plot of the modulation fraction (inset).  
}
\vspace{-0.15in}        
\label{density}
\end{figure}
 
For purposes of illustration, we model the velocity distribution function in the Galactic rest frame using the standard halo model (SHM):
 \es{SHMf}{
\tilde f (\V{v}) = \left\{ \begin{array}{ll}
{1 \over N_\esc } \left( {1 \over \pi v_0^2 } \right)^{3/2} e^{- \V{v}^2 / v_0^2 } \, \qquad &|\V{v}| < v_\esc \\
0 \, \qquad &\text{else} \,,
\end{array}
\right.
}
where $N_\esc$ is a normalization factor.  We take $v_0 \approx 220 \kms$ and the escape velocity to be $v_\esc \approx 550 \kms$~\cite{Smith:2006ym}.  While there is considerable debate about how well the SHM describes the tail of the distribution function, it appears to be a reasonable approximation at low speeds, which is most important for GF~\cite{Kuhlen:2009vh, Lisanti:2010qx}.

Consider the DM phase-space density in the lab-frame at time $t$, integrated over solid angle $\Omega$:
\es{angularInt}{
\rho f_v |_t \equiv \rho \, \int d \Omega \, v^2 f(\V{v}, t) \, .
}
Note that $\rho f_v|_t$ only depends on the speed $v$, as is relevant for experiments that are not sensitive to scattering direction.  For the SHM, the phase-space density changes only slightly in the Earth frame during the year and is  dominated by DM particles with speeds  
$\sim$$300 \kms$, as illustrated by the dashed black curve in Fig.~\ref{density}.  The relevant quantity for GF is $\left(\rho f_v|_{\text{Mar 1}} -  \rho f_v|_{\text{Sep 1}}\right)$, shown by the dotted red line, because it captures how many more DM particles impact Earth during the spring than the fall, for a given speed.  The two curves in Fig.~\ref{density} are individually normalized so their shapes can be compared.

The difference in the phase-space density between spring and fall is maximal for speeds \mbox{$\sim$200 km/s}.  This can be understood by looking at the fractional modulation between spring and fall, defined as $\left(\rho f_v|_\text{Mar 1} -  \rho f_v|_\text{Sep 1}\right) / \left(\rho f_v|_\text{Mar 1} + \rho f_v|_\text{Sep 1}\right)$; the difference between $\rho f_v|_\text{Mar 1}$ and $\rho f_v|_\text{Sep 1}$ is small enough that the denominator may be approximated as $2 \rho f_v$.
 We may understand heuristically how the fractional modulation changes as a function of $v$ through the following argument.  A particle traveling at speed $v$ has a crossing time $t \sim r_s / v$ near the gravitational well of the Sun as it travels from the Earth's location in September to the Earth's location in March.  During this time, the particle's trajectory is deflected a distance on the order of $(G M_\sun / r_s^2)\, t^2 \sim GM_\sun/v^2$.   Thus, the deflection as a fraction of $r_s$ is $ GM_\sun/(r_s v^2) \sim (v_\text{esc}^S / v)^2$. Here, $v_\text{esc}^S = \sqrt{2 G M_\sun / r_s} \approx 40 \kms$,  with $M_\sun$ the mass of the Sun, is the escape velocity to leave the Solar System from Earth's position.  The area $A$ subtended by the particle flow thus changes by $\delta\!A/A \sim (v_\text{esc}^S / v)^2$.  Because the flow rate of particles $\sim$$\rho A v$ is invariant in the Solar frame, and $v$ does not change much from one side of the Sun to the other for high-speed particles, the fractional change in the density scales as $\delta\!A/A$.  Therefore, 
the fractional modulation scales as $\sim$$(v_\text{esc}^S / v)^2$.  

The modulation fraction between spring and fall is shown in the inset of Fig.~\ref{density} and supports the intuition that GF has a more significant effect on lower-speed particles.  However, because there are more particles $\sim$$300 \kms$ in general, the difference in phase-space density between spring and fall is peaked at $\sim$$200 \kms$. 
 \begin{figure*}[tb]
\leavevmode
\begin{center}$
\begin{array}{cc}
\scalebox{.9}{\includegraphics{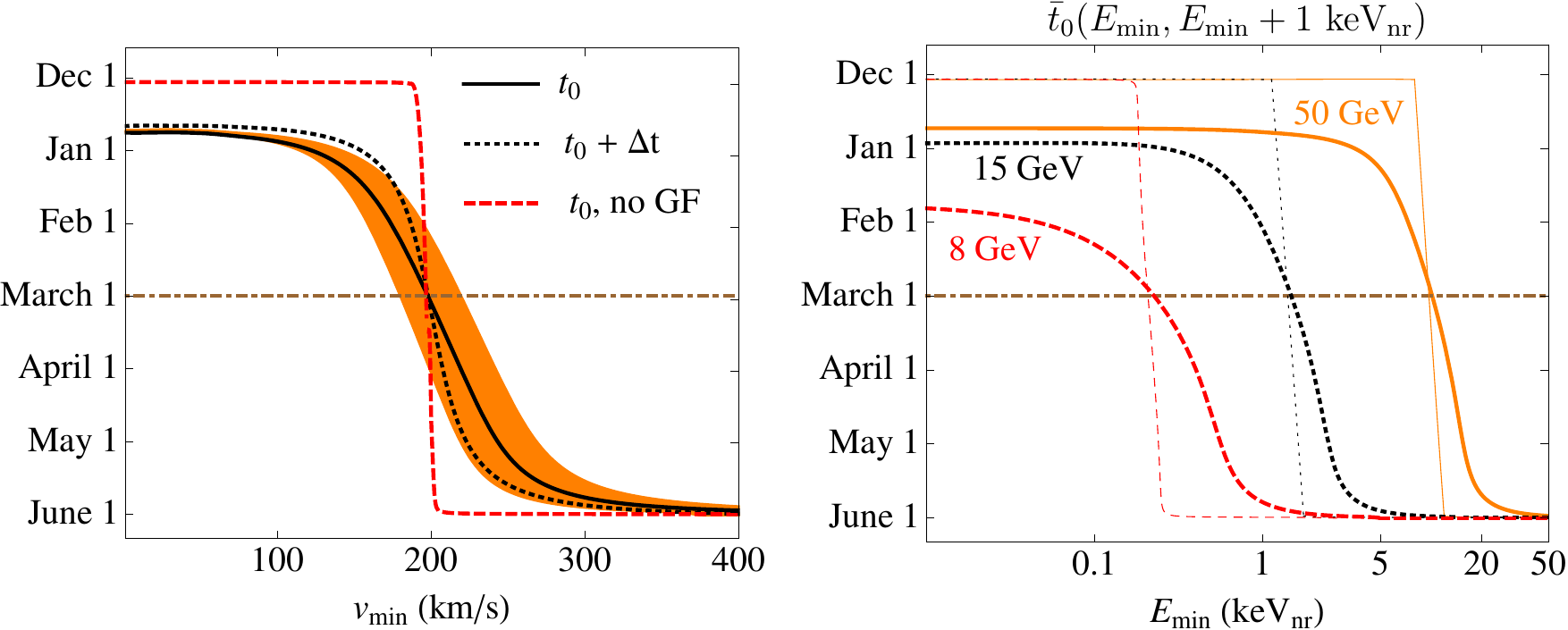}} \\
 \end{array}$
\end{center}
\vspace{-.50cm}
\caption{ (left) The time $t_0$ when the differential rate is maximized changes as a function of $v_\Min$.  Without GF, this time is around June 1 for $v_\Min \gtrsim 200 \kms$ and $\sim$half a year earlier for smaller $v_\Min$.  With GF, $t_0$ is still approximately June 1 at high $v_\Min$, but as $v_\Min$ decreases, GF becomes more significant, and $t_0$ ultimately asymptotes to a value $\approx 21$ days later than that expected with no GF.  The dot-dashed brown line marks March 1, which is the time when GF is maximal.  The orange region roughly accounts for the astrophysical error in $t_0$ by varying $v_0$ from $180$ to $260 \kms$.  (right)  The time $\bar t_0$ that maximizes the binned rate as a function of the minimal bin energy $E_\text{min}$ for DM masses 8, 15, and 50 GeV (thick lines).  The thin lines show the corresponding phases when GF is neglected.  We assume $1$ keV$_\text{nr}$ energy bins at a germanium detector; note that the shapes of these curves are highly sensitive to the bin size and target nucleus.  }
\vspace{-0.15in}
\label{PhaseRow}
\end{figure*}  

GF causes the total DM density to modulate by around 1.5\% between spring and fall.
Were GF the only physical effect causing the detection rate to modulate, we would expect the maximum of the rate to be \mbox{$\sim$March 1}.  Of course, this is not the case; the Earth's motion around the Sun causes the rate to modulate with an extremum around June 1, and the modulation fraction in this case is around 3\%.  At high $v_\Min$, the modulation due to the Earth's orbit dominates.  However, as $v_\Min$ approaches $\sim$$200 \kms$, GF becomes important.  For $v_\Min$ well below this scale, the expected maximum is around \mbox{December 1}, and GF causes it to shift towards \mbox{March 1} by around $(3 \, \, \text{months}) \times 1.5 / (3 + 1.5) \approx$ 1 month.  A more precise calculation shows that at low $v_\Min$, the maximum is shifted by around 21 days.
 
We now calculate the phase shift due to GF.
The time-dependence of the differential rate is captured by 
 \es{series}{
 {d R \over d E_\text{nr} } \approx A_0 + A_1 \, \cos \, \omega (t - t_0 - \Delta t) \,,
 }
where $A_1 \geq 0$, $t_0$ is the maximum of the rate, and $\Delta t$ is a possible phase shift that may arise from including the biannual and higher-frequency modes, which are not shown in~\eqref{series}.  
  For isotropic distributions, the biannual mode is suppressed relative to the mode $A_1$ by $\sim$$1/ 30$~\cite{Lee:2013xxa}.  This may be enhanced at high $v_\Min$, as it becomes increasingly sensitive to the Galactic escape velocity~\cite{Lee:2013xxa}.  GF also enhances the biannual mode, as captured by the phase $\Delta t$.     
When GF is neglected, $t_0 \approx$ June 1 at large $v_\Min$ and $\sim$half a year later at low $v_\Min$.
The transition occurs around $v_\Min \approx 200 \kms$, at which point $A_1$ has a zero, as is illustrated by the dashed red curve in the left panel of Fig.~\ref{PhaseRow}.

 GF may be accounted for by substituting the expression~\eqref{frelation} for $f(\V{v},t)$ into~\eqref{recoilR}.  Assuming the SHM, we compute the differential rate numerically, calculate the time $t_0$, and perform the harmonic expansion~\eqref{series} to find $\Delta t$.  The result for $t_0$ as a function of $v_\Min$ is shown in the left panel of Fig.~\ref{PhaseRow} (solid black).  The dotted black curve shows the phase shift $\Delta t$, which is most significant for $v_\Min$ near $200 \kms$.  The shaded orange region takes into account astrophysical uncertainties in the calculation of $t_0$ by varying $v_0$ from $180$ to $260 \kms$.  
Varying $\V{v_\sun}$~\cite{Nesti:2013uwa}  results in changes that are also contained within the shaded region.
 At low $v_\Min \ll 200 \kms$, we find that $t_0 \approx t_1 - 86$ days, where $t_1$ is the time of the vernal equinox.  This is $\sim$21 days later than the time one finds when neglecting GF, indicated by the dashed red line.
 
In practice, direct-detection experiments measure the differential rate in terms of $E_\text{nr}$ instead of $v_\Min$.  For elastic scattering, the two are related by \mbox{$v_\Min =  \sqrt{m_n E_\text{nr} / \left(2\, \mu^2\right)} $}, 
where $m_n$ is the mass of the nucleus and $\mu$ is the reduced mass of the DM-nucleus system.  Note that, for the same $E_\text{nr}$, a lighter DM particle has a larger $v_\Min$ than a heavier DM particle.  The consequence of this is that lighter DM requires lower energy thresholds to see the effect of GF.

Experiments typically measure scattering rates in finite energy bins, so the relevant quantity is
\es{RateInt}{
\bar R(E_\text{min}, E_\text{max}) = \int_{E_\text{min}}^{E_\text{max}} d E_\text{nr} \, {d R \over d E_\text{nr} } \, ,}
 for a bin with $E_\text{nr} = [E_\text{min}, E_\text{max}]$.  We use the notation $\bar t_0(E_\text{min}, E_\text{max})$ to refer to the time of maximal $\bar R(E_\text{min}, E_\text{max})$.  The right panel of Fig.~\ref{PhaseRow} shows $\bar t_0$ as a function of $E_\text{min}$, assuming a germanium target and \mbox{1 keV$_\text{nr}$} energy bins, for DM of mass $8$ GeV (dashed red), $15$ GeV (dotted black), and $50$ GeV (solid orange).  GF causes a phase shift of more than 10 days from {\mbox{June 1}} when $E_\text{min}$ is below $\sim$1, $3$, and 20 keV$_\text{nr}$ for 8, 15, and 50 GeV DM, respectively. The phase shift is particularly significant for $\gtrsim$ 15 GeV DM.  However, current advances in low-threshold technology  could make it possible to observe the effect of GF for an 8 GeV candidate.  For example, the projected threshold for SuperCDMS-lite is $\sim$0.8 keV$_\text{nr}$~\cite{Akerib:2004iz}.  GF can affect the phase of a light DM candidate at other low-threshold experiments, such as PICASSO~\cite{Archambault:2012pm} and DAMIC~\cite{Barreto:2011zu}.  
 For a xenon experiment, the phase shift is significant ($\gtrsim 10$ days) in the 4--5 keV$_\text{nr}$ energy bin for DM masses \mbox{$\gtrsim 20$ GeV}. 
 
The DAMA experiment, which uses a NaI(Tl) target, claims over 9$\sigma$ modulation with maximum at May 25$\pm$7 days 
in the energy range 2--6 keV$_\text{ee}$~\cite{Bernabei:2013xsa}.  From 2--5 keV$_\text{ee}$, the maximum is at May 21$\pm$6 days, while from 2--4 keV$_\text{ee}$, it is at May 15$\pm$6 days~\cite{Bernabei:2013xsa}.  The purported signal can correspond to $\sim$$10$ or 80 GeV DM, depending on whether the scattering off sodium or iodine dominates; both possibilities are in tension with null results from other experiments (see~\cite{Freese:2012xd} and references therein).  For a 10 GeV candidate, GF is not relevant in these energy bins and the expected maximum remains at June 1.  However, for 80 GeV DM, the maximum should be at $\sim$May 22, 21, and 19 in the energy ranges 2--6, 2--5, and 2--4 keV$_\text{ee}$, respectively.  A more dramatic effect should be observed in the 2--3 keV$_\text{ee}$ and 2--2.5 keV$_\text{ee}$ bins, where GF leads to a maximum at $\sim$May 10 and April 29, respectively.  

GF may be even more important if halo substructure with many slow-moving DM particles (in the Sun's reference frame) is present.  One such example is the dark disk (DD), which is thought to rotate in the same direction as the local standard of rest, but with a lag speed $\sim$$50 \kms$~\cite{Read:2008fh,Read:2009iv,Purcell:2009yp,Bruch:2008rx}.  In the low-$v_\Min$ limit, the DD may significantly alter the phase shift due to GF, even for relatively small DD density.  For example, if the DD makes up $10\%$ of the total DM,  $t_0$ changes at low $v_\Min$ from $\sim$December 24 (no DD) to 29.  
 At larger DD densities, $t_0$ asymptotes to around January 3.  Without accounting for GF, the DD causes the rate to be extremized $\sim$November 15~\cite{Lee:2013xxa}, so including GF corrects this phase shift by around 45 days.  For intermediate $v_\Min$, GF combined with the DD leads to an interesting signature in the modulation, which we do not investigate here.      
 
We have not included corrections from the GF effect due to the Earth's gravitational potential.  The velocity $v_\esc^E \approx 11 \kms$ to escape Earth's gravitational field from the surface is not significantly smaller than $v_\esc^S$, so one might expect it to be important.  However, the GF effect from the Earth does not modulate over the period of a year, and therefore does not significantly contribute to an annual modulation, as we checked numerically.  GF from the Earth should be significant for daily modulation, however, and GF from the moon should contribute to monthly modulation.  Along similar lines, DM particles trapped within the Solar system do not contribute significantly to annual modulation because (i) their density is small compared to that of unbound DM~\cite{Peter:2009mi,Peter:2009mm}, and (ii) the velocity distribution of the bound DM should not modulate over the period of a year.  

To summarize, the gravitational potential of the Sun focuses incoming, unbound DM particles, leading to an observable phase shift in the modulation signal.  For DM with $v_\text{min} \lesssim 200 \kms$, the signal is maximal \mbox{$\sim$21} days later than would be expected with no focusing.  These results have important implications for DM masses \mbox{$\gtrsim 20$--50 GeV}, depending on target mass.  They also are relevant for light DM in low-threshold experiments.

\vspace{-0.19in} 
\section*{Acknowledgments}
\vspace{-0.15in}
We thank J. Hall for helpful discussions.  ML is supported by a Simons Postdoctoral Research Fellowship.  BRS is supported by the NSF grant PHY-0756966.

\onecolumngrid
\vspace{0.3in}
\twocolumngrid
\def\bibsection{} 
\bibliographystyle{apsrev}
\bibliography{DMmod}

\end{document}